\documentstyle[prb,aps]{revtex}

\twocolumn

\begin{document}

\title{The Surface
Energy of a Bounded Electron Gas: Analysis of the Accuracy of the
Local-Density Approximation via Ab Initio Self-Consistent-Field 
Calculations\\}
\author{J. M. Pitarke$^1$ and A. G. Eguiluz$^{2,3}$} 
\address{$^1$Materia Kondentsatuaren Fisika Saila, Zientzi
Fakultatea, Euskal Herriko  Unibertsitatea,\\ 644 Posta kutxatila, 48080 Bilbo,
Basque Country, Spain\\
$^2$Department of Physics and Astronomy, The University of Tennessee, Knoxville,
TN 37996-1200\\
$^3$Solid State Division, Oak Ridge National Laboratory, Oak Ridge,TN
37831-6032}
\date\today
\maketitle

\begin{abstract}
We report an ab initio evaluation of the surface energy of a simple
metal, performed via a coupling-constant integration over the dynamical
density-response function.  The rapid rate of change of the electron density at
the surface  is treated exactly.  Long-range correlations are treated
self-consistently in the random-phase approximation; short range correlations
are included  in time-dependent local density-functional  theory.  Our results
provide a numerical measure of the error introduced by the usual local-density
approximation; this error is found to be small.
\end{abstract}


\narrowtext

Since the pioneering work of Lang and Kohn\cite{Lang70}, the 
calculation of the surface energy of a metal has been the
subject of long-standing interest.  These authors were the 
first to include the crucial effects of exchange and 
correlation self-consistently within the local-density 
approximation (LDA) of density-functional theory (DFT) 
\cite{Hohenberg}.  Lang and Kohn also discussed the effect
of the crystal lattice, whose full inclusion within the LDA poses no 
difficulties of principle these days.  By contrast, the question of the 
impact of non-local Coulomb correlations, and their interplay 
with the strong charge inhomogeneity at the surface, has remained unsettled 
over the years\cite{Harris}$^-$\cite{Perdews}.

Recent calculations have rekindled the controversy on 
the question of the quality of the LDA surface energy.  A 
many-body scheme which starts from a physically-motivated 
model of the correlated ground state, and treats the interactions
via a Fermi hypernetted-chain 
approximation (FHNCA)\cite{Kohn86}, has yielded surface energies
which are significantly higher than the LDA results of 
Lang and Kohn\cite{Lang70}.  By contrast, the surface 
energies obtained in density-functional calculations\cite{Zhang} 
based on the use of the Langreth-Mehl non-local functional 
\cite{Langreth81} are much closer to the LDA result.  Finally, although very
recent quantum Monte Carlo (QMC)calculations\cite{Ceperley92} agree with the
latter non-local density functional results\cite{Zhang} for high densities
($r_s\le 2.07$), they agree with the FHNCA for lower densities ($r_s\ge 3.25$).

The purpose of this paper is to establish, in a controlled way, the impact of
non-locality on the surface energy of an electron gas.  To this end we carry
out non-local and local calculations within exactly the same conditions, i.e.,
we consider the LDA as a special case of the general non-local formalism based
on a coupling-constant integration over the dynamical density-response
function.First, the effects of non-local 
correlations are investigated fully self-consistently within 
a well-defined many-body framework, the random-phase
approximation (RPA)\cite{Pines}. Our side-by-side calculation, in which the same
diagram is used to generate the local and
non-local surface energies, shows that the LDA is quite
accurate over the entire density domain appropriate to  
metals ($r_s=2-6$). Of course, the absolute values of our
RPA energies cannot be expected to be more reliable than, say, the QMC surface
energies. However, the significance of our results lies in the elucidation of the
{\it difference} between non-local and local surface
energies. We also explore the impact of short-range correlations by invoking a
time-dependent extension of local density-functional theory
(TDLDA)\cite{Soven}. The TDLDA vertex introduces an element of arbitrariness, since
it contains no dynamical effects. Thus, again, we are less interested in the absolute
value of the TDLDA surface energy than we are in its {\it difference} with its local
counterpart; again, we evaluate this difference in an unambiguous way. Our TDLDA
results support the conclusion drawn from our RPA results that the error introduced
by the LDA is not large.

The ground-state energy of an interacting electron system can 
be written as a functional of
the electron number density $n({\bf r})$\cite{Hohenberg},
\begin{equation}
E[n]=E_k[n]+E_{es}[n]+E_{xc}[n],
\end{equation}
where $E_k[n]$ is the kinetic energy of
a non-interacting system with the same density and $E_{es}[n]$
is the Hartree electrostatic energy.  The exchange-correlation (XC) energy,
$E_{xc}[n]$, can be obtained from a
coupling constant-integration over the interaction energy.  We follow Langreth
and Perdew\cite{Langreth75} and choose the coupling constant 
such that the density is maintained at its fully-interacting 
value while the electron-electron interaction strength is varied 
from $\lambda=0$ to $\lambda=1$.  The
fluctuation-dissipation theorem then leads us to the result that
\begin{eqnarray}
E_{xc}[n]=&&{e^2\over 2}\int_0^1{\rm d}\lambda\int{\rm d}^3{\bf r}\int{\rm
d}^3{\bf r}' {1\over|{\bf r}-{\bf
r}'|}\cr&&\times\left[-{\hbar\over\pi}\int_0^\infty{\rm d}\omega
\chi_\lambda({\bf r},{\bf r}';{\rm
i}\omega)-n({\bf r})\delta({\bf r}-{\bf r}')\right],
\end{eqnarray}
where $\chi_\lambda({\bf r},{\bf r}';\omega)$ is the
density-response function.

If the Coulomb correlations are ignored altogether, Eq. (2)
reduces to the expression
\begin{eqnarray}
E_x[n]=&&{e^2\over 2}\int{\rm d}^3{\bf r}\int{\rm d}^3{\bf r}'
{1\over|{\bf r}-{\bf r}'|}\cr&&\times\left[-{\hbar\over\pi}\int_0^\infty
{\rm d}\omega\chi^0({\bf r},{\bf r}';{\rm
i}\omega)-n({\bf r})\delta({\bf r}-{\bf r}')\right],
\end{eqnarray}
where 
\begin{eqnarray}
\chi^0({\bf r},{\bf
r}';\omega)=&&2\sum_{i,j}{\theta(E_F-E_i)-\theta(E_F-E_j)\over
E_i-E_j-\hbar(\omega+{\rm i}\eta)}\cr &&\times\psi_i({\bf r})\psi_j^*({\bf
r})\psi_j({\bf r}')\psi_i^*({\bf r}')
\end{eqnarray}
is the density-response function for non-interacting electrons.  For 
{\it inhomogeneous} systems Eq. (3) coincides with the Fock exchange 
energy only if the wave functions $\psi_i({\bf r})$ are the 
solutions of the non-local Hartree-Fock equation.

Actually, Eq. (4) can be interpreted in a more general context 
as giving the density-response to an appropriate mean field 
set up by the dynamical polarization of the Fermi sea.  In the 
particular case of the RPA, the 
single-particle wave functions $\psi_i({\bf r})$ entering Eq. (4) are
strictly the self-consistent eigenfunctions of the one-electron Hartree 
Hamiltonian. In time-dependent density-functional theory\cite{Runge}, the
"non-interacting" electrons in question are described by the solutions
of the time-dependent counterpart of the Kohn-Sham
equation\cite{Hohenberg}; in usual practice, these amplitudes 
are approximated by standard LDA wave functions \cite{Gross}.  Both 
approaches to the evaluation of the polarizability $\chi^0({\bf r},{\bf r}';
\omega)$ 
will be considered in our numerical study of the surface
energy of a simple metal.

In both RPA and TDLDA the response function satisfies the integral
equation\cite{note1} 
\begin{eqnarray}
\chi_\lambda&&({\bf r},{\bf r}';\omega)=\chi^0({\bf r},{\bf r}';\omega)\cr&&
+\int{\rm d}{\bf r}_1\int{\rm d}{\bf r}_2\chi^0({\bf r},{\bf r}_1;\omega)\lambda
V({\bf r}_1,{\bf r}_2)
\chi_\lambda({\bf r}_2,{\bf r}';\omega).
\end{eqnarray}
In the RPA, the effective electron-electron interaction $V({\bf r}_1,{\bf r}_2)$
is just the bare Coulomb potential. In TDLDA,
\begin{equation}
V({\bf r}_1,{\bf r}_2)=
{e^2\over|{\bf r}_1-{\bf r}_2|}+{{\rm d} V_{xc}({\bf r}_1)\over
{\rm d} n({\bf r}_1)}\delta({\bf r}_1-{\bf r}_2),
\end{equation}
$V_{xc}$ being the local XC potential. The 
combination of Eqs. (2) and (5) defines 
either the TDLDA or the RPA exchange-correlation energy, depending 
on whether or not the Coulomb interaction is "dressed" according to Eq. (6).  
This dressing corresponds to the inclusion of short-range correlations, which 
are ignored in RPA.

We consider a jellium slab of thickness $a$ and density
$\overline n_+=q_F^3/3\pi^2$, where 
$q_F=(9\pi/4)^{1/3}/(r_sa_0)$ is the Fermi
wave vector, $r_s$ is the Wigner-Seitz radius, and $a_0$ is the Bohr radius. 
The slab is translationally invariant in the plane of 
the surface, which is assumed to be
normal to the $z$ axis.  Thus the single-particle 
wave functions are of the form
\begin{equation}
\psi_i({\bf r})={1\over\sqrt A}\phi_i(z){\rm e}^{{\rm
i}{\bf q}_\parallel\cdot{\bf r}_\parallel},
\end{equation}
where ${\bf q}_\parallel$ is a wave vector parallel to the surface and $A$ is 
the normalization area.  The wave functions $\phi_i(z)$ describe motion normal 
to the surface, and are obtained self-consistently with the
effective one-electron potential,
$V_{eff}(z)$.  In the RPA, $V_{eff}$ consists of just the Hartree 
potential $V_{es}(z)$;  in TDLDA, 
\begin{equation} V_{eff}(z)=V_{es}(z)+V_{xc}(z).
\end{equation}
For reference, we recall that,
in the simplest non-self-consistent microscopic model of the surface,
the infinite-barrier model (IBM), $V_{eff}(z)$ is replaced by an infinite square
barrier, and the functions $\phi_i(z)$ are simply sines.
 
A solution of Eq. (5) for a self-consistent description of the surface potential 
has been given some time ago\cite{Eguiluz}.  We assume that
$n(z)$ vanishes at a distance $z_0$ from either jellium edge, and expand
the wave functions $\phi_i(z)$ in a Fourier sine series;  $z_0$
is chosen sufficiently large for the physical results to be insensitive to the 
precise value employed.  We introduce a double-cosine Fourier 
representation for the density response
function, and also for the Coulomb potential and the Dirac delta function 
entering Eq. (2)\cite{Pitarke96}.  The use of this representation allows us 
to perform analytically the integrals 
in Eq. (2) involving the coordinate normal to the surface; the integrals over 
parallel-momentum transfers, over 
energy transfers, and over the coupling constant, are 
performed numerically.  The total energy given by Eq.
(1) is evaluated in a similar way.  Subtracting from the total 
energy the corresponding result for a
homogeneous electron gas of density $\overline n_+$, $E^H\left[\overline
n_+\right]$, we obtain the surface energy
\begin{equation}
\sigma={1\over 2A}\left\{E\left[n(z)\right]-E^H\left[\overline n_+\right]
\right\}.
\end{equation}

The LDA is obtained from the above non-local formalism by replacing 
the response function entering Eq. (2) by its counterpart for a homogeneous  
electron gas with the local value of the density.  This replacement 
leads us to the result that\cite{Pitarke96}
\begin{equation}
E_{xc}^{LDA}\left[n\right]={1\over a}\int_0^{a+2z_0}{\rm d}
z E_{xc}^H\left[n(z)\right].
\end{equation}
Equation (10), with $E_{xc}^H\left[n(z)\right]$ 
evaluated on the basis of (homogeneous electron gas-) RPA 
and TDLDA density-response functions, calculated for the 
local value of the density $n(z)$ obtained self-consistently with 
Hartree and LDA effective potentials, yields our RPA- and TDLDA-based 
LDA surface energies.
We will consider these local results together with the LDA 
results of Lang and Kohn\cite{Lang70} $-$which we 
also obtain from Eq. (10) through the use of the Wigner
interpolation formula for $E_{xc}^H\left[n(z)\right]$ \cite{Wigner}. 

Great care was exercised to ensure that our slab calculations are a faithful 
representation of the surface energy of a semi-infinite medium.  This issue is 
important, in view of the subtle cancellations which exist between the various 
contributions to the surface energy; furthermore, these contributions are
oscillatory functions of the slab width $a$. (The amplitude of the
oscillations decays approximately linearly with $a$; their period equals
$\lambda_F/2$, $\lambda_F=2\pi/q_F$ being the Fermi wavelength.)

For each value of $r_s$ we have actually 
considered three different values of $a$. One such value, $a_n$, is
the threshold  width for which the $n$-th subband for $z$-motion is first
occupied; for this width the surface energy is a local minimum. The other two
values of the slab width, 
$a_n^-=a_n-\lambda_F/4$, and $a_n^+=a_n+\lambda_F/4$, correspond to the
two local maxima about the minimum. Utilizing the relation\cite{Pitarke96}
\begin{equation}
\sigma={\sigma_n^-+\sigma_n+\sigma_n^+\over 3},
\end{equation}
we are able to extrapolate our calculated surface energy
to the infinite-width limit. This procedure was first tested, with values of $n$
up to $200$, for the IBM, for which analytical insight is possible by virtue of
the simple nature of the one-electron wave functions\cite{Pitarke96}. The
results presented below correspond to slabs with $n=12$, for which $a\approx
5-6\lambda_F$, depending on
$r_s$. Based on this procedure, we estimate that the numerical error
introduced by our slab simulations corresponds to one unit in the last digit of
all the entries in Table I. (We remark that our results were found to be
insensitive to the precise value of the number of sines, $s_{max}$, kept in the
expansion of the wave functions $\phi_i(z)$, for
$s_{max}\ge 280$.)

The key results of our work can be readily grasped from Fig. 1, in 
which we show the surface energy as a function of $r_s$. 
Consider first the RPA. The reasons for the significance of our RPA calculations
are: (i) the effects of long-range correlations are included fully
self-consistently with the electron density profile (which, we recall, is
evaluated in the Hartree approximation); (ii) the non-local and local
calculations are carried out within one and the same density-response framework;
(iii) this framework is devoid of any ambiguities in the treatment of the
many-body problem.  {\it It is clear from Fig. 1 that the local RPA surface
energy differs little from its non-local counterpart over the entire metallic
range of densities.}

Next, we consider the effects of the short-range correlations built into the XC 
potential, $V_{xc}$.  In the full TDLDA treatment, this effect is included
in both  the one-electron potential of Eq. (8) and the electron-electron
interaction of Eq. (6). (We evaluate $V_{xc}$ with use of the Perdew and Zunger
parametrization\cite{PerdewZ}.)  Overall, the main impact of the inclusion of XC
is via Eq. (8), which, through self-consistency, yields a more abrupt
electron density  profile at the surface, relative to the Hartree profile. This
leads to the large lowering of the surface energy, relative to the RPA, which we
observe in Fig. 1.  For completeness, in Fig. 1 we show both the full TDLDA
(solid line), and the result obtained upon including XC in Eq. (8) but not in
Eq. (6) (dashed line);  clearly, the impact of the XC
vertex is smaller than the effect of the inclusion of XC 
in the electron density profile. 

As was the
case above with the RPA calculations, {\it the difference between non-local}
(solid line) {\it and local} (dotted line) {\it TDLDA surface energies is
relatively small}. More specifically, the error introduced by the LDA is of the
order of $50\%$ smaller than the error one would impute to the local
approximation on the basis of the non-local FHNCA results \cite{Kohn86} (open
circles in Fig. 1), particularly in the crucial high-density region ($r_s\approx
2$).

It is apparent from Fig. 1 that our non-local TDLDA
surface energies agree well, {\it for all densities}, with those obtained 
by Zhang et al.\cite{Zhang} using the
non-local Langreth-Mehl\cite{Langreth81} XC functional. 
By contrast, while the surface energy obtained via QMC techniques\cite{Ceperley92}
is close to either set of results for $r_s=2.07$,  for lower densities the QMC surface
energies are appreciably larger; in fact, 
they are close to the FHNCA values.

It is interesting to note that our TDLDA-based LDA surface energies agree closely with
the LDA calculations of Acioli and 
Ceperley\cite{Ceperley92}. (Our results can also be reproduced 
from Eq. (10) with use of the Perdew and Zunger
parametrization of $E_{xc}^H[n(z)]$\cite{PerdewZ}.)  Thus, 
while from the QMC results of Ref.\onlinecite{Ceperley92} one would conclude that the
error introduced by the LDA is, for $r_s> 3.25$, rather significant, this is not what
transpires from our results $-$most particularly, from our RPA calculations.

We note that the difference between our TDLDA-based LDA results and
the local Lang-Kohn surface energies (dashed-dotted-dotted-dotted line) is simply a
consequence of the use by these authors\cite{note2}of the Wigner formula for
correlation\cite{Wigner}.  The much larger difference between Lang-Kohn
surface energies and our RPA-based LDA results reflects, again, the impact of XC
on the electron density profile at the surface.

If the surface energy is obtained from Eq. (3) (not Eq. (2)), in conjunction with
an {\it exchange-only} correction to the Hartree barrier in Eq. (8),  we obtain
non-local surface energies which are close to the Hartree-Fock results reported
by Krotscheck et al.\cite{Kohn86}  By contrast, the non-local {\it correlation}
contribution to the surface energy reported in Ref.\onlinecite{Kohn86} is
significantly higher than our ab initio {\it correlation} surface energy; as a
result, the FHNCA surface energies are much higher than our non-local TDLDA
values, as shown in Fig. 1.  The FHNCA and RPA surface energies turn out to be
quite close, at low densities, because of compensation between the effect of XC
in the effective one-electron potential, which is absent in the RPA, and the
very large FHNCA non-local correlation surface energy.

%Table I lists XC and total surface energies obtained in RPA and TDLDA.
%Local XC contributions to the surface energy are systematically smaller
%than their non-local counterparts, a result also obtained in the
%IBM\cite{Lang75,Langreth75}. In the IBM the electron density profile at
%the surface is, when $z$ is measured in units of $\lambda_F$, independent of
%$r_s$; thus the percentage difference between local and non-local XC surface
%energies is the same ($6.5\%$) for all values of $r_s$.  In the self-consistent
%RPA and TDLDA, the density profile  becomes more abrupt as $r_s$ increases %(with
%$z$ measured in  units of $\lambda_F$); thus the percentage error of local XC
%surface energies increases monotonically with
%$r_s$.

%It is clear from Table I (and Fig. 1) that the local surface energy
%deviates more from its non-local TDLDA counterpart than is the case in the RPA. 
%However, since the use of a static vertex in Eq. (6) is an uncontrolled
%approximation, it is not possible to state unequivocally that the
%potentially-superior TDLDA results are a more reliable arbiter of the impact of
%non-locality than the RPA results.

In conclusion, we have presented ab initio calculations of the surface energy
of a bounded electron gas. The unambiguous nature of the
comparison of local- vs. non-local surface energies made possible by our
self-consistent RPA calculations, leads us to the conclusion that the LDA does,
within the RPA, work. We have also evaluated the {\it difference} between non-local
and local TDLDA surface energies, and the results so obtained support the conclusion
that the error introduced by the LDA is, within the TDLDA, not large, over the whole
range of electron densities appropriate to metals. Further progress in the quantitative
ab initio evaluation of this {\it difference} requires improvements in
the treatment of dynamical many-body correlations beyond the scope of Eq. (6).

J. M. P. acknowledges partial support by the Spanish Ministerio 
de Educaci\'on y Ciencia, the University of the Basque
Country and the Basque Unibertsitate eta Ikerketa Saila. A. G. E.
acknowledges support from National Science Foundation Grant No. DMR-9634502 
and from the National Energy Research Supercomputer Center. ORNL is managed by
Lockheed Martin Energy Research Corp. for the U. S. DOE under contract
DE-AC05-96OR22464.

\narrowtext
\begin{table}
\caption{Exchange-correlation ($\sigma_{xc}$) and total ($\sigma$)
non-local surface energies obtained in RPA and TDLDA, and their local
counterparts. Units are $erg/cm^2$.} 
%, $\sigma_{xc}^{LDA}$ and \sigma^{LDA}$
\begin{tabular}{lrrrrrrrr}
&\multicolumn{4}{c}{RPA}&\multicolumn{4}{c}{TDLDA}\\
$r_s$&$\sigma_{xc}$&$\sigma_{xc}^{LDA}$&$\sigma$
&$\sigma^{LDA}$&$\sigma_{xc}$&$\sigma_{xc}^{LDA}$&$\sigma$
&$\sigma^{LDA}$\\
\tableline

2.0&4657&4583&-126&-200&3533&3353&-686&-866\\
2.07&4154&4080&73&-1&3125&2959&-446&-612\\
3.0&1203&1175&477&449&840&763&301&224\\
4.0&467&454&281&268&295&261&198&164\\
5.0&226&219&164&157&130&111&117&98\\
6.0&125&121&100&96&65&54&71&60\\
\end{tabular}
\label{table1}
\end{table}

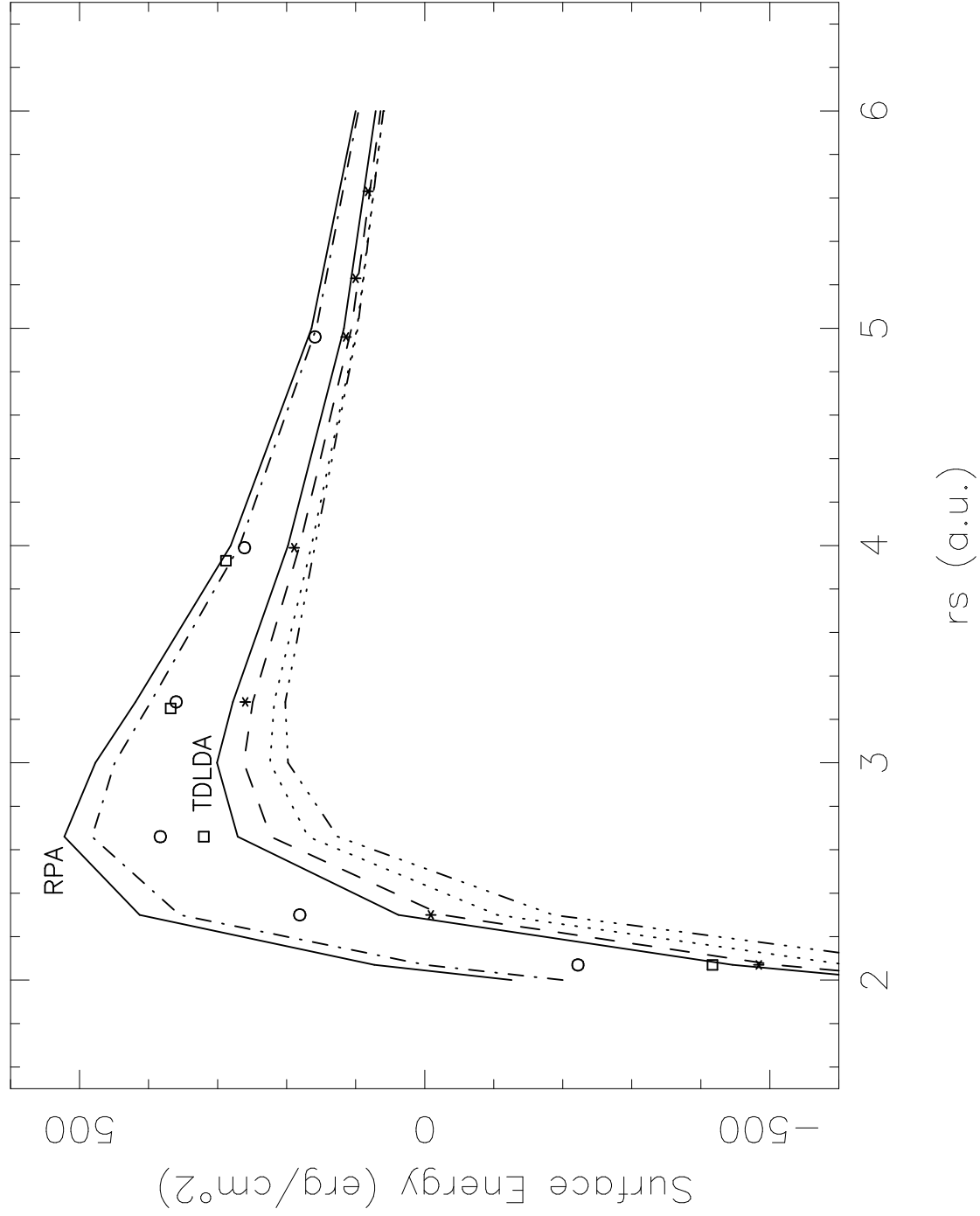
\begin{figure}
\caption{Non-local RPA and TDLDA surface energies (solid lines), as
functions of \protect$r_s$\protect. The dashed
line is the TDLDA result obtained upon excluding the XC vertex from Eq. (6). 
Dashed-dotted and dotted lines represent the local RPA and TDLDA surface
energies, respectively.  The dashed-dotted-dotted-dotted line represents the
Lang-Kohn surface energy of Ref.\protect\onlinecite{Lang70}\protect. Open
circles, stars and open squares are results taken from
Refs.\protect\onlinecite{Kohn86}\protect\protect$\,$\protect,
\protect\onlinecite{Zhang}\protect\protect$\,$\protect and
\protect\onlinecite{Ceperley92}\protect, respectively.}
\end{figure}


\begin{references}
\bibitem[23]{note2} Differences between our TDLDA-based LDA results and the
Lang-Kohn surface energies also appear as a result of these authors evaluating
the XC potential of Eq. (8) by using the Wigner parametrization. However, we
have found that the various contributions to both local and non-local
exchange-correlation surface
energies are rather insensitive to the choice of the correlation potential of
Eq. (8), and that total surface energies are even more insensitive to this
choice.
\end{references}
\end{document}